# SECURITY ISSUES IN THE OPTIMIZED LINK STATE ROUTING PROTOCOL VERSION 2 (OLSRv2)

Ulrich Herberg[1] and Thomas Clausen[2]

Hipercom@LIX, Ecole Polytechnique, France
[1]ulrich@herberg.name, [2]thomas@thomasclausen.org

## ABSTRACT

*Mobile Ad hoc NETworks (MANETs) are leaving the confines of research laboratories, to find place in real-world deployments. Outside specialized domains (military, vehicular, etc.), city-wide community-networks are emerging, connecting regular Internet users with each other, and with the Internet, via MANETs. Growing to encompass more than a handful of "trusted participants", the question of preserving the MANET network connectivity, even when faced with careless or malicious participants, arises, and must be addressed. A first step towards protecting a MANET is to analyze the vulnerabilities of the routing protocol, managing the connectivity. By understanding how the algorithms of the routing protocol operate, and how these can be exploited by those with ill intent, countermeasures can be developed, readying MANETs for wider deployment and use.*

*This paper takes an abstract look at the algorithms that constitute the Optimized Link State Routing Protocol version 2 (OLSRv2), and identifies for each protocol element the possible vulnerabilities and attacks – in a certain way, provides a "cookbook" for how to best attack an operational OLSRv2 network, or for how to proceed with developing protective countermeasures against these attacks.*

## 1 INTRODUCTION

OLSRv2 (the Optimized Link State Routing Protocol version 2) [1], [2], [3], [4], [5] is a successor to the widely deployed OLSR [6] routing protocol for MANETs (Mobile Ad hoc NETworks). OLSRv2 retains the same basic algorithms as its predecessor, however offers various improvements, e.g. a modular and flexible architecture allowing extensions, such as for security, to be developed as add-ons to the basic protocol.

The developments reflected in OLSRv2 have been motivated by increased real-world deployment experiences, e.g. from networks such as FunkFeuer [7], and the requirements presented for continued successful operation of these networks. With participation in such networks increasing (the FunkFeuer community network has, e.g., roughly 400 individual participants), operating with the assumption, that participants can be "trusted" to behave in a non-destructive way, is utopia. Taking the Internet as an example, as participation in the network increases and becomes more diverse, more efforts are required to preserve the integrity and operation of the network. Most SMTP-servers were, e.g., initially available for use by all and sundry on the Internet – with an increased populace on the Internet, attacks and abuses caused the recommended practice is today to require authentication and accounting for users of such SMTP servers [8].

A first step towards hardening against attacks disrupting the connectivity of a network, is to understand the vulnerabilities of routing protocol, managing the connectivity. This paper therefore analyzes OLSRv2, to understand its inherent vulnerabilities and resiliences. The authors do not claim completeness of the analysis, but hope that the identified attacks, as presented, form a meaningful starting-point for developing secured OLSRv2 networks.

## 1.1 OLSRv2 Overview

OLSRv2 contains three basic processes: Neighborhood Discovery, MPR Flooding and Link State Advertisements, described in the below with sufficient details for elaborating the analysis in latter sections of this paper.

### 1.1.1 Neighborhood Discovery

Neighborhood Discovery is the process, whereby each router discovers the routers which are in direct communication range of itself (1-hop neighbors), and detects with which of these it can establish bi-directional communication. Each router sends HELLOs, listing the identifiers of all the routers from which it has recently received a HELLO, as well as the "status" of the link (heard, verified bi-directional). A router $a$ receiving a HELLO from a neighbor $b$, in which $b$ indicates to have recently received a HELLO from $a$, considers the link $a$-$b$ to be bi-directional. As $b$ lists identifiers of all its neighbors in its HELLO, $a$ learns the "neighbors of its neighbors" (2-hop neighbors) through this process. HELLOs are sent periodically, however certain events may trigger non-periodic HELLOs.

### 1.1.2 MPR Flooding

MPR Flooding is the process whereby each router is able to, efficiently, conduct network-wide broadcasts. Each router designates, from among its bi-directional neighbors, a subset (MPR set) such that a message transmitted by the router and relayed by the MPR set is received by all its 2-hop neighbors. MPR selection is encoded in outgoing HELLOs. Routers may express, in their HELLO messages, their "willingness" to be selected as MPR, which is taken into consideration for the MPR calculation, and which is useful for example when an OLSRv2 network is "planned". The set of routers having selected a given router as MPR is the MPR-selector-set of that router. A study of the MPR flooding algorithm can be found in [9].

### 1.1.3 Link State Advertisement

Link State Advertisement is the process whereby routers are determining which link state information to advertise through the network. Each router must advertise, at least, all links between itself and its MPR-selector-set, in order to allow all routers to calculate shortest paths. Such link state advertisements are carried in TCs, broadcast through the network using the MPR flooding process described above. As a router selects MPRs only from among bi-directional neighbors, links advertised in TC are also bi-directional and routing paths calculated by OLSRv2 contain only bi-directional links. TCs are sent periodically, however certain events may trigger non-periodic TCs.

## 1.2 Link State Vulnerability Taxonomy

Proper functioning of OLSRv2 assumes that (i) each router can acquire and maintain a topology map, accurately reflecting the effective network topology; and (ii) that the network converges, *i.e.* that all routers in the network will have sufficiently identical topology maps. An OLSRv2 network can be disrupted by breaking either of these assumptions, specifically (a) routers may be prevented from acquiring a topology map of the network; (b) routers may acquire a topology map, which does not reflect the effective network topology; and (c) two or more routers may acquire inconsistent topology maps.

### 1.3 OLSRv2 Attack Vectors

Besides "radio jamming", attacks on OLSRv2 consist of a malicious router injecting "correctly looking, but invalid, control traffic" (TCs, HELLOs) into the network. A malicious router can either (a) lie about itself (its ID, its willingness to serve as MPR), henceforth *Identity Spoofing* or (b) lie about its relationship to other routers (pretend existence of links to other routers), henceforth *Link Spoofing*. Such attacks will *in-fine* cause disruption in the Link State Advertisement process, through targeting the MPR Flooding mechanism, or by causing incorrect link state information to be included in TCs, causing routers to have incomplete, inaccurate or inconsistent topology maps. In a different class of attacks, a malicious router injects control traffic, tuned to cause an *in-router resource exhaustion*, *e.g.* by causing the algorithms calculating routing tables to be invoked continuously, preventing the internal state of the router from converging.

### 1.4 Paper Outline

The remainder of this paper is organized as follows: section 2 presents related work to this paper. Section 3, 4, 5 and 6 each represents a class of disruptive attacks against OLSRv2, detailing a number of attacks in each class. Section 7 summarizes inherent resilience, as observed in OLSRv2, and the paper is concluded in section 8.

## 2 RELATED WORK

A number of articles has analyzed security properties and vulnerabilities of routing protocols in MANETs ([10], [11], [12], [13]). These papers identify resources of MANET routing protocols that are potentially vulnerable to attacks, and propose several attacks against these resources, as well as counter-measures against such attacks. [14], [15], [16] present a more detailed security analysis of the OLSR routing protocol (in its first version as specified in [6]). However, we believe that our paper presents a much more detailed analysis of the successor of OLSR, by taking an abstract look at each of the algorithms that constitute OLSRv2 and by identifying for each protocol element the possible vulnerabilities and attacks. Moreover, while the main basic algorithms of OLSR have been integrated into OLSRv2, some of the presented vulnerabilities in this paper are specific to OLSRv2.

## 3 TOPOLOGY MAP ACQUISITION

Topology Map Acquisition relates to the ability for a given router in the network to acquire a representation of the network connectivity. A router, unable to acquire a topology map, is incapable of calculating routing paths and of forwarding data. Topology map acquisition can be hindered by (a) TCs to not being delivered to (all) routers in the network, such as what happens in case of flooding disruption, or (b) in case of jamming of the communication channel.

### 3.1 Flooding Disruption

MPR selection (section 1.1.2) uses information about a router's 1-hop and 2-hop neighborhood, assuming that (i) this information is accurate, and (ii) all 1-hop neighbors are equally apt as MPR. Thus, a malicious router seeking to attack the MPR Flooding process will seek to manipulate the 1-hop and 2-hop neighborhood information in a router such as to cause the MPR selection to fail.

#### 3.1.1 Flooding Disruption due to Identity Spoofing

Figure 1(a) illustrates a network in which the malicious router (gray circle) spoofs the identity of *b*, *i.e. a* receives HELLOs from two routers, both pretending to be *b*. As HELLOs are additive, and with the malicious router *X* not advertising any neighbors, the topological view of the 1-hop and 2-hop neighborhood of *a* is unaffected by the presence of *X*: *a*'s MPR selection will function correctly by selecting (if using a greedy algorithm) *d*.

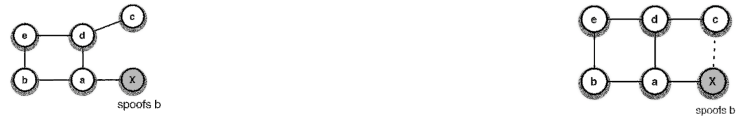

(a) The gray malicious router spoofs *b*

(b) X spoofs *b* and advertises a link to *c*

Figure 1: Identity Spoofing: flooding attack: 1-hop address duplication.

Figure 1(b) illustrates a network in which the malicious router *X* (gray circle) spoofs the identity of *b*. In this example, a link (the dotted line) between *X* and *c* is correctly detected and advertised by *X*. Router *a* will receive HELLOs indicating that links exist from *b* to both *e* and *c*, thereby rendering *b* a candidate MPR on par with *d*.

If *X* does not forward flooded traffic (*i.e.* does not accept MPR selection), its presence entails a flooding disruption: selecting *b* over *d* renders *c* unreachable by flooded traffic. In order to increase the likelihood that the malicious *X* is selected, it may set its willingness to 7 (max), ensuring that it is always selected and the routers so covered are not further considered in the MPR selection algorithm.

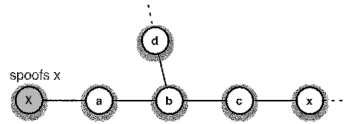

Figure 2: Identity Spoofing: flooding attack: 2-hop address duplication.

Figure 2 illustrates a network in which the malicious router *X* (gray circle) spoofs the identity of *x*, *i.e. a* and *c* both receive HELLOs from a router pretending to be *x*. From the point of view of *b*, it appears as if *a* and *c* hve the same neighbor set, hence either is a suitable choice as MPR. Assuming that *b* selects *a* as MPR, *c* will not relay flooded traffic and thus the legitimate (white) *x* (and routers to the "right" of *x*) will not receive flooded traffic.

In order to maximize the impact of the disruption, the malicious router may simultaneously "spoof" multiple identities: by overhearing control traffic for a while, the malicious router may attempt to learn the identities of neighbors of *c* and spoof these – and, in addition, assume one additional identity (possibly not otherwise present in the network). A way of achieving this is to simply have *X* overhear all TCs, and spoof all identities of all routers in the network (possibly excluding *a*). Router *b* will learn through the HELLOs of *a* that all these identities are 2-hop neighbors of *a*. As the set of identities spoofed by the malicious *X* is a superset of the neighbors of *c*, this will cause selection of *a* as MPR, and consequently that *c* is not selected.

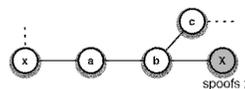

Figure 3: Identity Spoofing: flooding attack: 1 and 2-hop address duplication.

Figure 3 illustrates a network in which the malicious router *X* (gray circle) spoofs the identity of *x*, *i.e. a* and *b* both receive HELLOs from a router pretending to be *x*. Router *b* will therefore not select *a* as MPR as all the 2-hop neighbors reachable via *a* are already reachable directly in one hop. As a consequence, the white *x*, and any routers "to the left" of it, will not receive flooded control traffic from *b* or transited via *b* from *e.g. c*.

### 3.1.2 Flooding Disruption due to Link Spoofing

Figure 4(a) illustrates a network, in which the malicious router *X* spoofs links to the non-existing *c*, *i.e. a* receives HELLOs from *X*, pretending the existence of a link between *X* and *c*. This forces *a* to select *X* as MPR – whereas it otherwise would not need to select any MPRs. In this simple example, this does no harm as such.

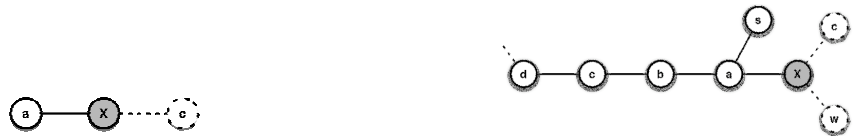

(a) Link Spoofing effect on MPR Selection    (b) Flooding disruption due to link spoofing

Figure 4: Link Spoofing: Flooding Disruption

Figure 4(b) illustrates a network, in which the malicious *X* spoofs links to the existing *c*, as well as to a non-existing *w*. Router *a* receives HELLO from *X* reporting links to *c* and *w*, and from *b* reporting a link to *c* only. Unless if *b* has a advertised a willingness of 7, this will cause *a* to select *X* as its only MPR, as *X* presumably covers all 2-hop neighbors of *a* (*i.e.* the real neighbors of *a* as well as the imaginary *w*).

The consequence is that as *a* will not select *b* as MPR, *b* will not relay flooded messages received from *a*. Thus, the network to the left of *b* (starting with *c*) will not receive any flooded messages from or transiting *a*, such as a message originating from *s* and transiting through *a*.

## 3.2  Radio Jamming

Radio jamming is an attack in which legitimate access to the communications channel between routers is forcefully hindered by a malicious device. The classic example hereof is where a powerful transmitter is generating "white noise" over the communications channel where the network interfaces of the routers would otherwise operate, effectively preventing these router interfaces from successfully receiving transmissions from each other. While this can happen on all network interface and channel types, wireless networks are especially vulnerable to such; commercial WiFi "jammers" are, for example, readily available [17].

The consequence of such jamming is that the router interfaces, which are so "jammed", are unable to receive routing protocol control traffic, and so are unable to participate in the network. A router where all its network interfaces are victim to "jamming" is, effectively, unable to acquire a topology map of the network and, so, is disconnected from the network.

It can be observed that a router with multiple network interfaces accessing different communications channels, and where not all communications channels are jammed, may still be able to participate in a network via links over these non-jammed interfaces.

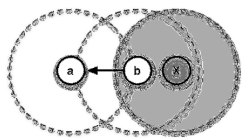

Figure 5: Radio Jamming

It can also be observed that while direct jamming affects reception, it may (depending on which lower layers L1/L2 are employed) not affect transmission. Thus, and as illustrated in figure 5, *a*, may receive transmissions from *b*, the latter of which is otherwise "jammed" by *X*, which prevents receptions in the grayed area.

The Neighborhood Discovery mechanism of OLSRv2 identifies uni- and bidirectionality of links, and only bi-directional links are advertised and used for routing path calculations. OLSRv2 has, thus, by virtue of this detection and use of only bi-directional links, some resilience to jamming: while the jammed routers are unable to acquire and maintain a topology map of the network, the jammed routers appear as simply "disconnected" to the un-jammed part of the network – which is able to both maintain accurate and consistent topology maps.

## 3.3 Attack on Jittering

OLSRv2 incorporates jittering: a random, but bounded, delay on outgoing control traffic. This may be necessary when link layers (such as 802.11 [18]) are used, which do not guarantee collision-free delivery of frames, and where jitter can reduce the probability of collisions of frames on lower layers is [1].

In OLSRv2, TC forwarding is jittered by a value between 0 and `MAX_JITTER`. In figure 6, a router receives three packets, each containing a TC to be forwarded. For each of these, the scheduled retransmission time is calculated as "now plus jitter", illustrated by the horizontal arrows.

In order to reduce the number of transmissions, when a control message is due for transmission, OLSRv2 piggybags all queued messages into a single transmission. Thus, if a malicious router sends many TCs within a very short time interval, the jitter time of the attacked router tends to 0. This renders jittering ineffective and can lead to collisions.

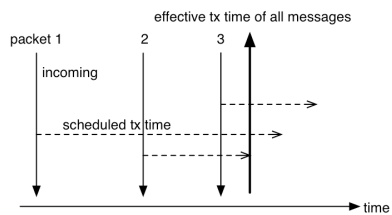

Figure 6: Jittering Attack

## 3.4 Hop-count and Hop-limit Attacks

The hop-count and hop-limit fields are the only parts of a TC that are modified when forwarding. A malicious router can modify either of these when, when forwarding TCs.

### 3.4.1 Modifying the Hop Limit

A malicious router can decrease the hop limit when forwarding a TC. This will reduce the scope of forwarding the message, and may lead to some routers in the network not receiving that TC. Note that this is not necessarily the same as *not* relaying the message (i.e. setting the hop limit to 0), as illustrated in figure 7.

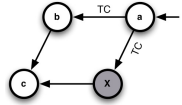

Figure 7: Hop limit attack

A TC arrives at and is forwarded by *a*, such that it is received by both *b* and the malicious *X*. *X* can forward the TC without any delay (including without jitter) such that its transmissions arrives before that of *b* at *c*. Before forwarding, it significantly reduces the hop limit of the message. Router *c* receives the TC, processes (and forwards) it, and marks it as already received - causing it to discard further copies received from *b*. Thus, if the TC is forwarded by *c*, it has a very low hop limit and will not reach the whole network.

### 3.4.2 Modifying the Hop Count

A malicious router can modify the hop count when forwarding a TC. This may have two consequences: (i) if the hop count is set to the maximum value, then the TC will be forwarded no further by, or (ii) artificially manipulating the hop count may affect the validity time as calculated by recipients, when using distance-dependent validity times as defined in [3] (*e.g.* as part of a fish-eye extension to OLSR2 [19]).

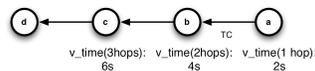

Figure 8: Different validity times based on the distance in hops

In figure 8, *a* sends a TC with a validity time of two seconds for neighbors that are one hop away, four seconds for routers in a two-hop distance and six seconds in a three-hop distance. If *c* is a malicious router and modifies the hop count (say, by decreasing it to 0), then *d* will calculate the validity time of received information to two seconds – after which it expires unless refreshed. If TCs from *a* are sent less frequently than that up to 3 hops, this causes links advertised in such TCs to be only intermittently available to *d*.

## 4 EFFECTIVE TOPOLOGY

Link-state protocols assume that each router can acquire an accurate topology map, reflecting the *effective network topology*. This implies that the routing protocol, through its message exchange, identifies a path from a source to a destination, and this path is valid for forwarding data traffic. If an attacker disturbs the correct protocol behavior, the perceived topology map of a router can permanently differ from the effective topology.

Considering the example in figure 9(a), which illustrates the topology map as acquired by *s*. This topology map indicates that the routing protocol has identified that for *s*, a path exists to *d* via *b*, which it therefore assumes can be used for transmitting data. If, effectively, *b* does not forward data traffic from *s*, then the topology map in *s* does not accurately reflect the effective network topology. Rather, the effective network topology from the point of view of *s* would be as indicated in figure 9(b): *d* is not part of the network reachable from router *s*.

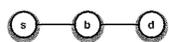          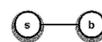

(a) Perceived topology by *s*                          (b) Effective topology

Figure 9: Incorrect Data Traffic Forwarding

### 4.1 Incorrect Forwarding

OLSRv2 routers exchange information using link-local transmissions (link-local multicast or limited broadcast) for their control messages, with the routing process in each router retransmitting received messages destined for network-wide diffusion. Thus, if the operating system in a router is not configured to enable forwarding, this will not affect the operating of the routing protocol, or the topology map acquired by the routing protocol. It will, however, cause a discrepancy between the effective topology and the topology map, as indicated in figure 9(a) and figure 9(b). This situation is not hypothetical. A common error seen when deploying OLSRv2 based networks using Linux-based computers as router is to neglect enabling IP forwarding.

### 4.2 Wormholes

A wormhole, depicted in the example in figure 10, may be established between two collaborating devices, connected by an out-of-band channel; these devices send traffic through the "tunnel" to their alter-ego, which "replays" the traffic. Thus, d and s appear as-if direct neighbors and reachable from each other in 1 hop through the tunnel, with the path through the MANET being 100 hops long.

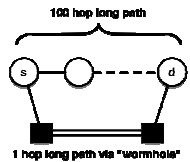

Figure 10: Wormhole between 2 collaborating devices not participating in the routing protocol.

The consequences of such a wormhole in the network depends on the detailed behavior of the wormhole. If the wormhole relays only control traffic, but not data traffic, the same considerations as in section 4.1 applies. If, however, the wormhole relays all traffic, control and data alike, it is connectivity-wise identical to a usable link – and the routing protocol will correctly generate a topology map reflecting the effective network topology. The efficiency of the topology so obtained depends on (i) the wormhole characteristics, (ii) how the wormhole presents itself and (iii) how paths are calculated.

Assuming that paths are calculated with unit-cost for all links, including the "link" presented by the wormhole: if the real characteristics of the wormhole are *as-if* it was a path of more than 100 hops (*e.g.* with respect to delay, bandwidth, ....), then the presence of the wormhole results in a degradation in performance as compared to using the non-wormhole path. Conversely, if the "link" presented by the wormhole has better characteristics, the wormhole results in improved performance.

If paths are calculated using non-unit-costs for all links, and if the cost of the "link" presented by the wormhole correctly represents the actual cost (*e.g.* if the cost is established through measurements across the wormhole), then the wormhole may in the worst case cause no degradation in performance, in the best case improve performance by offering a better path. If the cost of the "link" presented by the wormhole is misrepresented, then the same considerations as for unit-cost links apply.

An additional consideration with regards to wormholes is, that such may present topologically attractive paths for the network – however it may be undesirable to have data traffic transit such a path: an attacker could, by virtue of introducing a wormhole, acquire the ability to record and inspect transiting data traffic.

### 4.3 Sequence Number Attacks

OLSRv2 uses two different sequence numbers in TCs, to (i) avoid processing and forwarding the same message more than once (Message Sequence Number), and (ii) to ensure that old information, arriving late due to *e.g.* long paths or other delays, is not allowed to overwrite fresher information (Advertised Neighbor Sequence Number – ANSN).

#### 4.3.1 Message Sequence Number

An attack may consist of a malicious router spoofing the identity of another router in the network, and transmitting a large number of TCs, each with different Message Sequence Numbers. Subsequent TCs with the same sequence numbers, originating from the router whose identity was spoofed, would thence be ignored, until eventually information concerning these "spoofed" TCs expires.

#### 4.3.2 Advertised Neighbor Sequence Number (ANSN)

An attack may consist of a malicious router spoofing the identity of another router in the network, and transmitting a single TC, with an ANSN significantly larger than that which was last used by the legitimate router. Routers will retain this larger ANSN as "the most fresh information" and discard subsequent TCs with lower sequence numbers as being "old".

### 4.4 Message Timing Attacks

In OLSRv2, each control message may contain explicit "validity time" and "interval time", identifying the duration for which information in that control message should be considered valid until discarded, and the time until the next control message should be expected [3].

#### 4.4.1 Interval Time Attack

A use of the expected interval between two successive HELLOs is for determining the link quality in Neighbor Discovery process, as described in [6]: if messages are not received with the expected intervals (*e.g.* a certain fraction of messages are missing), then this may be used to exclude a link from being considered as useful, even if (some) bi-directional communication has been verified. If a malicious $X$ spoofs the identity of an existing $a$, and sends HELLOs indicating a very low interval time, $b$ receiving this HELLO will expect the following HELLO to arrive within the interval time indicated – or otherwise, decrease the link quality for the link $a$-$b$. Thus, $X$ may cause $b$'s estimate of the link quality for the link $a$-$b$ to fall below the limit, where it is no longer considered as useful and, thus, not used.

#### 4.4.2 Validity Time Attack

A similar attack – with respect to the interval time attack – uses the validity time included in HELLO and TCs. The validity time defines how long the information contained in the message should be considered as valid. After this time, the receiving router must consider the message content to no longer be valid (unless repeated in a later message) [3]. A malicious router, $X$, can spoof the identity of a $a$ and send a HELLO using a very low validity time (*e.g.* 1 ms). $b$, receiving this, will discard the information upon expiration of that interval, *i.e.* a link $a$-$b$ will be "torn down" by $X$.

### 4.5 Indirect Jamming

Indirect Jamming is an attack in which a malicious router is, by its actions, causing legitimate routers to generate inordinate amounts of control traffic, thereby increasing both channel occupation and the overhead incurred in each router for processing this control traffic. This control traffic will be originated from legitimate routers, thus to the wider network, the malicious device may remain undetected.

The general mechanism whereby a malicious device can cause indirect jamming is for it to participate in the protocol by generating plausible control traffic, and to tune this control traffic to in turn trigger receiving routers to generate additional traffic. For OLSRv2, such an indirect attack can be directed at, respectively, the Neighborhood Discovery mechanism and the Link State Advertisement mechanism.

### 4.5.1 Indirect Jamming: Neighborhood Discovery

An indirect jamming attack on the Neighborhood Discovery process is illustrated in figure 11.

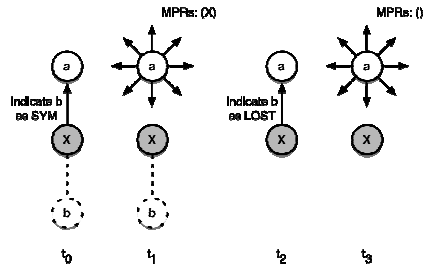

Figure 11: Indirect Jamming in Neighborhood Discovery

A malicious router, $X$, advertises in a HELLO that it has as link to $b$, with status SYM ($t_0$). This will cause $a$, upon receiving this HELLO, to consider $b$ as a 2-hop neighbor, and recalculate its MPR set – selecting $X$ as MPR. This MPR selection is signaled by $a$ in a subsequent HELLO ($t_1$). Upon receipt of this HELLO from $a$, $X$ advertises in a HELLO that the link to $b$ is LOST ($t_2$). This will cause $a$, upon receiving this HELLO, to no longer consider $b$ as a 2-hop neighbor, and recalculate its MPR set accordingly, *i.e.* to no longer contain $X$. This new MPR set in $a$ is signaled in a subsequent HELLO ($t_3$). Upon $X$ having received this HELLO from $a$, it may repeat the cycle, alternating advertising the link $X$-$b$ as LOST and SYM.

In order to maximize the impact of the disruption caused by this attack, $X$ should ensure that the router, to which it alternatively advertises a link as SYM or LOST, is not otherwise present in the 2-hop neighborhood – for example by advertising a router not otherwise present in the network. That way, all neighbors receiving a HELLO from $X$ will select $X$ as MPR. A way of accomplishing this is to have $X$ learn all identities in the network by overhearing all TCs – and selecting (spoofing) an identity not already present. $X$ will indicate its willingness to be non-zero (thus, accepting being selected as MPR) and participate in the Neighborhood Discovery procedure – and may ignore all other protocol operations, while still remaining effective as an attacker.

An easier version of this attack is to have $X$ simply be present in the network, and participate in the Neighborhood Discovery procedure. Without spoofing a link to another router, $X$ alternates its willingness as advertised in successive HELLO transmissions between zero (will never be selected as MPR) and 7 (will always be selected as MPR). The impact of this version of the attack is as above: MPR set recalculation and advertisement by neighbors of the $X$.

The basic Neighborhood Discovery process of OLSRv2 employs periodic message emissions, and by this attack it can be ensured that for each message exchange between $X$ and $a$, the MPR set in $a$ is recalculated. As calculation of an optimal MPR set is known to be NP-hard [20], this alone may cause internal resource exhaustion in $a$.

If the routers in the network have "triggered HELLOs" enabled, and that such are triggered by MPR set updates (as suggested in section 9 in [5]) this attack may also cause an increased

HELLO frequency. A minimum message interval (typically much smaller than the regular peri-
odic message interval) is imposed, to rate-limit worst-case message emissions. This attack can
cause the HELLO interval to, permanently, become equal to the minimum message interval. [5]
proposes that default that the minimum HELLO interval be 1/4 HELLO interval.

Indirect Jamming of the Neighborhood Discovery process by a malicious router can thus have
two effects: to cause increased frequency of HELLO generation and transmission by neighbors
of the malicious router, *i.e.* up to two hops away from the malicious router, and to cause addi-
tional MPR set calculation in the routers which are neighbors of a malicious router.

### 4.5.2 Indirect Jamming: Link State Advertisement

The most efficient Indirect Jamming attack in OLSRv2 is to target control traffic, destined for
network-wide diffusion. This is illustrated in figure 12.

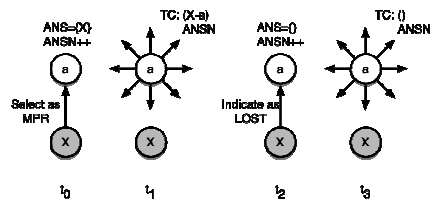

Figure 12: Indirect Jamming in Link State Advertisement

The malicious $X$ selects the $a$ as MPR ($t_0$) in a HELLO. This causes $X$ to appear as MPR selector
for $a$ and, consequently, $a$ sets $X$ to be advertised in its "Neighbor Set" and increments the asso-
ciated "Advertised Neighbor Sequence Number" (ANSN). $a$ must, then, advertise the link be-
tween itself and $X$ in subsequent outgoing TCs ($t_1$), also including the ANSN in such TCs. Upon
$X$ having received this TC, it declares the link between itself and $a$ as no longer valid ($t_2$) in a
HELLO (indicating the link to $a$ as LOST). Since only symmetric links are advertised by
OLSRv2 routers, $a$ will upon receipt hereof remove $X$ from the set of advertised neighbors and
increment the ANSN. $a$ will then in subsequent TCs advertise the remaining set of advertised
neighbors (*i.e.* with $X$ removed) and the corresponding ANSN ($t_3$). Upon $X$ having received this
information in another TC from $a$, it may repeat this cycle, alternating advertising the link $a$-$X$ as
"LOST" and as "MPR".

Routers receiving a TC will parse and process this message, specifically updating their topology
map as a consequence of successful receipt. If the ANSN between two successive TCs from the
same router has incremented, then the topology has changed and routing tables are to be recalcu-
lated. This is a potentially computationally costly operation [21].

A malicious router may chose to conduct this attack against all its neighbors, thus attaining
maximum disruptive impact on the network with relatively little overhead of its own: other than
participating in the Neighborhood Discovery procedure, the malicious router will monitor TCs
generated by its neighbors and alternate the advertised status for each such neighbor, between
"MPR" and "LOST". The malicious router will indicate its willingness to be zero (thus, avoid
being selected as MPR) and may ignore all other protocol operations, while still remaining effec-
tive as an attacker.

The basic operation of OLSRv2 employs periodic message emissions, and by this attack it can
be ensured that each message will entail routing table recalculation in all routers in the network.

If the routers in the network have "triggered TCs" enabled, this attack may also cause an in-
creased TC frequency. Triggered TCs are intended to allow a (stable) network to have relatively low TC emission frequencies, yet still allow link breakage or link emergence to be advertised through the network rapidly. A minimum message interval (typically much smaller than the regular periodic message interval) is imposed, to rate-limit worst-case message emissions. This attack can cause the TC interval to, permanently, become equal to the minimum message interval. [5] proposes as default that the minimum TC interval be 1/4 TC interval.

Indirect Jamming by a malicious router can thus have two effects: it may cause increased frequency of TC generation and transmission, and it will cause additional routing table recalculation in all routers in the network.

## 5 INCONSISTENT TOPOLOGY

Inconsistent topology maps can occur by a malicious router employing either of identity spoofing or link spoofing for conducting an attack against an OLSRv2 network.

### 5.1 Identity spoofing

Identity spoofing can be employed by a malicious router via the Neighborhood Discovery process and via the Link State Advertisement process; either of which causing inconsistent topology maps in routers in the network.

#### 5.1.1 Inconsistent Topology Maps due to Neighborhood Discovery

Considering the network in figure 13, two routers in far ends of the network both present themselves under the same identity – as $x$. The routers adjacent to these two routers ($a$ and $w$ in figure 13) both perceive $x$ as a direct neighbor, which will be reflected in the neighbor tables and routing tables of these two routers. Thus, the first consequence is, that traffic destined for $x$ from $a$ and $w$, respectively, will be delivered to different routers. As the Neighborhood Discovery procedure also provides topological information up to two hops away, this is also true for traffic destined for $x$ from $b$ and $v$, respectively.

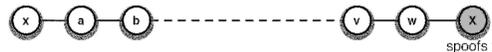

Figure 13: Identity Spoofing: a router (gray circle) assumes the identity of router $x$

Assuming unit-cost links, the distance to $x$ from $a$ and $w$, as produced by the Neighborhood Discovery procedure, is 1 hop. The distance to $x$ from $b$ and $v$, as produced by the Neighborhood Discovery procedure, is 2 hops. As these distances are shorter than (or equal to) the path lengths obtained via the Link State Advertisement procedure, they will therefore be preferred by $a$, $b$, $v$ and $w$, over those acquired via the Link State Advertisement procedure for when calculating routing tables. Thus, if the gray router $X$ in figure 13 is the one spoofing the identity of the white router $x$, then any traffic from or transiting through $w$ and destined for $x$ will be delivered to the gray router $X$ instead of to the white $x$.

This has as impact that a router spoofing the identity of another router, and by simply participating in the Neighborhood Discovery procedure, will be able to alter the topology maps in routers up to 2 hops away, and thereby (i) attract the traffic from or transiting through routers up to two hops away, which is otherwise destined for the router whose identity is being spoofed; and (ii) prevent traffic from or transiting through routers up to two hops away, which is otherwise destined for the router whose identity is being spoofed, from reaching the intended destination.

Strategic placement of a malicious router spoofing the identity of another router (or other routers) in the network, and simply participating only in the Neighborhood Discovery process, can thereby efficiently disrupt network connectivity. First, overhearing TCs will allow the router to "learn" sufficient information describing the network topology to develop an attack strategy which has maximum disruptive impact. Second, by participating only in the Neighborhood Discovery procedure (*i.e.* by advertising its willingness as zero, and by not selecting MPRs), and by carefully selecting the identities to spoof, the malicious router can remain difficult to detect.

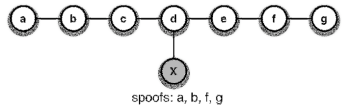

Figure 14: Identity Spoofing: maximizing disruptive impact while minimizing risk of detection.

Consider the example in figure 14: *X* overhears and learns the network topology. In order to minimize the risk of detection, it elects to not select any MPRs (thereby no Link State Advertisements are sent, advertising its presence in the network) and advertises its willingness as zero (thereby it is not selected as MPR and thus not required to send Link State Advertisements). *X* also elects to spoof the identity of *a*, *b*, *f* and *g* only. As *X* does not participate in the Link State Advertisement process, its presence is known only to *c*, *d* and *e*, *i.e.* the routers whose identity it spoofs will not receive control messages allowing them to detect that these identities are *also* advertised elsewhere in the network. Traffic transiting *d*, from either side, to destinations *a*, *b*, *f* and *g* will, rather than being forwarded to the intended destination, be delivered to *X*. Traffic transiting *c* and with *b* as destination will be delivered to the intended router *b*. Traffic transiting *c* and with *a* as destination may be delivered to the intended router *a* via *b* or to *X* via *d* – as the paths will be of equal length.

In figure 14, *c* is the only router which will receive control traffic indicating two topologic locations of the identities *a*, *b*. However, especially in a wireless environment, this is not in and by itself unusual: a valid link might indeed exist between *a* and *d* as well as between *b* and *d*, *e.g.* through another wireless channel. Thus, the topology as perceived by *c* and *e* does not appear "improbable".

If the network grows to the left of *a* or to the right or *g*, all *X* has to do to continue disrupting the network is to "learn" the identities of the routers beyond *a* and *g* and also spoof the identities of these. In general, for maximum disruptive impact and minimum visibility, the malicious router would select to spoof the identities of all routers which are topologically 3 hops or more away from itself.

Identity spoofing by a malicious router, strictly participating only in the Neighborhood Discovery process, thus, creates a situation wherein two or more routers have substantially inconsistent topology maps: traffic for an identified destination is, depending on where in the network it appears, delivered to different routers.

### 5.1.2 Inconsistent Topology Maps due to Link State Advertisements

An inconsistent topology map may also occur when the malicious router takes part in the Link State Advertisement (LSA) procedure, by selecting a neighbor as MPR, which in turn advertises the spoofed identities of the malicious router. This attack will alter the topology maps all routers of the network.

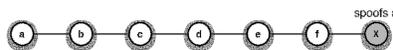

Figure 15: Identity Spoofing: router *X* spoofs *a*, leading to a wrongly perceived topology

In figure 15, *X* spoofs the address of *a*. If *X* selects *f* as MPR, all routers in the network will be informed about the link *f-a* by the TCs originating from *f*. Assuming that (the real) *a* selects *b* as MPR, the link *b-a* will also be advertised in the network, resulting in a perceived topology as depicted in figure 16.

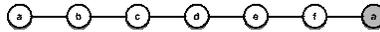

Figure 16: Identity Spoofing: the gray malicious router spoofs the identity of router *a*

When calculating paths, *b* and *c* will calculate paths to *a* via *b*, as illustrated in figure 17(a); for these routers, the shortest path to *a* is via *b*. *e* and *f* will calculate paths to *a* via *f*, as illustrated in figure 17(b); for these routers, the shortest path to *a* is via the malicious router *X*, and these are thus disconnected from the real *a*. *d* will have a choice: the path calculated to *a* via *b* is of the same length as the path via the malicious router *X*, as illustrated in figure 17(b).

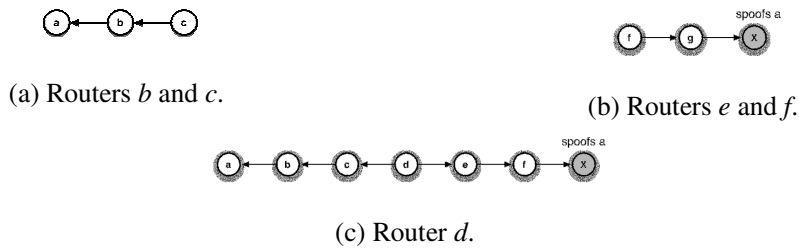

(a) Routers *b* and *c*.

(b) Routers *e* and *f*.

(c) Router *d*.

Figure 17: Routing paths towards *a*, as calculated by the different routers in the network

In general, the following observations can be made:

• The network will be split in two, with those routers closer to *b* than to *X* reaching *a*, whereas those routers closer to *X* than to *b* will be unable to reach *a*.

• Routers beyond *b* will be unable to detect this identity spoofing.

The identity spoofing attack via the Link State Advertisement procedure has a higher impact than the attack described in section 5.1.1, since it alters the topology maps of all routers in the network, and not only in the 2-hop neighborhood. However, the attack is easier to detect by other routers in the network. Since the malicious router is advertised in the whole network, routers whose identities are spoofed by the malicious router can detect the attack. For example, when *a* receives a TC from *f* advertising the link *f-a*, it can deduce that some entity is injecting incorrect Link State information as it does not have *f* as one of its direct neighbors.

As the malicious router *X* does not itself send the TCs, but rather, by virtue of MPR selection, ensures that the addresses it spoofs are advertised in TCs from its MPR selector *f*, the attack may be difficult to counter: simply ignoring TCs that originate from *f* may also suppress the link state information for other, legitimate, MPR selectors of *f*.

Identity spoofing by a malicious router, participating in the Link State Advertisement process by selecting MPRs only, thus, creates a situation wherein two or more routers have substantially inconsistent topology maps: traffic for an identified destination is, depending on where in the network it appears, delivered to different routers.

## 5.2 Link Spoofing

Link Spoofing is a situation in which a router advertises non-existing links to another router (possibly not present in the network). Essentially, TCs and HELLOs both advertise links to di-

175

rect neighbor routers, with the difference being the scope of the advertisement. Thus, link spoofing consists of a malicious router, reporting that it has as as neighbors routers which are, either, not present in the network, or which are effectively not neighbors of the malicious router.

It can be noted that a situation similar to Link Spoofing may occur temporarily in an OLSRv2 network without malicious routers: if $a$ was, but is no more, a neighbor of $b$, then $a$ may still be advertising a link to $b$ for the duration of the time it takes for the the Neighborhood Discovery process to determine this changed neighborhood.

In the context of this paper, Link Spoofing refers to a persistent situation where a malicious router intentionally advertises links to other routers, for which it is not a direct neighbor.

### 5.2.1 Inconsistent Topology Maps due to Neighborhood Discovery

Returning to figure 4(b), MPR selection serves to identify which routers are to advertise which links in the network as part of the Link State Advertisement process.OLSRv2 stipulates that a router must, as a minimum, advertise links between itself and its MPR selectors, *i.e.* links between itself and the routers which have selected it as MPR. A router is not required to advertise other links. Thus, in the example network in figure 4(b) with $a$ selecting the malicious router $X$ as its sole MPR, only $X$ is expected to advertise links to $a$. $s$ selects $a$ as its MPR, thus $a$ is expected to advertise the link $a$-$S$. $s$, then, expects $a$ to have selected suitable MPRs for the MPR flooding process to succeed in network-wide diffusion of the advertisement of the link $a$-$s$.

The topology maps acquired by the various other routers in this example are:

• **Routers $a$ and $b$** will, due to the Neighborhood Discovery process providing topological information up to 2 hops away, acquire an accurate Topology Map. For $a$ this is exactly corresponding to the network in figure 4(b). For $b$ this may or may not contain the dotted routers $c$ and $w$, depending on whether $X$ generates Link State Advertisements (see section 5.2.2).

• **Router $c$** will perceive a topology map as illustrated in figure 18(a): the link state advertisements from $a$ are not forwarded by $b$, hence the existence of $s$ and the link $a$-$s$ is not known beyond $b$; the same is true for a link state advertisement from $X$, should it participate in the link state advertisement process. The link $b$-$a$, and the existence of $a$ is known to $b$ only through the Neighborhood Discovery process.

• **Routers $d$ and beyond** will receive a Topology Map as illustrated in figure 18(b).

• **Router $s$** will acquire an accurate Topology Map corresponding to the network in figure 4(b). This may or may not contain the dotted routers $c$ and $w$, depending on if router $X$ generates Link State Advertisements (see section 5.2.2).

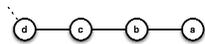

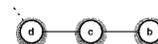

(a) Inconsistent Topology Map in $c$.　　　(b) Inconsistent Topology Map in $d$

Figure 18: Perceived Topology Maps with $X$ performing Link Spoofing

In order to maximize the impact of the disruption of Link Spoofing in the Neighborhood Discovery process, the malicious router may simultaneously "spoof" links to multiple routers: by overhearing control traffic "for a while", $X$ may attempt to learn the identities of 2-hop neighbors of $a$ and spoof these – and, in addition, assume at least one additional identity (possibly not otherwise present in the network). A way of achieving this is to simply have the malicious $X$ overhear all TCs, and spoof links to all identities of all routers in the network, plus one identity not otherwise present in the network. As the set of links spoofed by $X$ is thus a superset of the 2-hop links as seen from $a$, $a$ will select $X$ as its sole MPR.

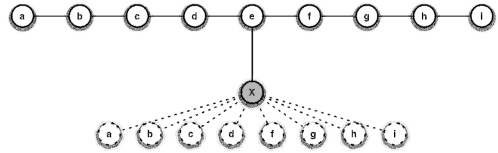

Figure 19: Link Spoofing: Malicious router *X* spoofs links to all routers in the network except *e*

Symmetric to figure 14, figure 19 illustrates a network with *X* is positioned in the middle. If *X* advertises links to *a*, *b*, *c*, *d*, *f*, *g*, *h* and *i* in the Neighborhood Discovery process, these identities as spoofed by *X* are visible only to *e* as 2-hop neighbors. *e* may detect that no link to *X* is advertised by its own 1-hop neighbor routers *d* and *f*. Thus, to avoid such detection by *e*, *X* should avoid spoofing links to routers advertised as 1-hop neighbors by *e*, *i.e.* advertise in its HELLOS only spoofed links to *a*, *b*, *c*, *g*, *h* and *i*, as illustrated in figure 20.

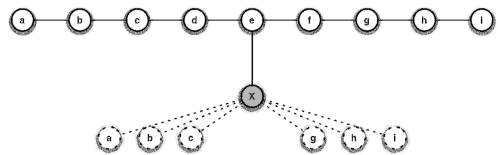

Figure 20: Link Spoofing: *X* does not advertise spoofed links to routers at most 2 hops away

The impact of this attack is that:

• *X* will appear as the most attractive candidate MPR for *e*, by virtue of spoofing links to all other 2-hop neighbors of *e* – and then some. Thus, absent *d* or *f* indicating a willingness of 7, *X* will be selected as the sole MPR of *e*.

• No routers, other than *X*, will be requested to send TCs, advertising links to *a*.

• No routers, other than *X*, will be requested to forward flooded traffic originating in or transiting through *a*.

### 5.2.2 Inconsistent Topology Maps due to Link State Advertisements

Figure 21 illustrates a network, in which the malicious router *X* spoofs links to the existing router *a* by participating in the Link State Advertisement process and including this non-existing link in its advertisements.

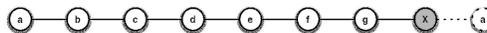

Figure 21: Link Spoofing: The malicious router *X* advertises a spoofed link to *a* in its TCs.

As TCs are flooded through the network, all routers will receive and record information describing a link *X-a* in this link state information. If *a* has selected router *b* as MPR, *a* will likewise flood this link state information through the network, thus all routers will receive and record information describing a link *b-a*.

When calculating routing paths, *b*, *c* and *d* will calculate paths to *a* via *b*, as illustrated in figure 22(a); for these routers, the shortest path to *a* is via *b*. *f* and *g* will calculate paths to *a* via *X*, as illustrated in figure 22(b); for these routers, the shortest path to *a* is via *X*, and these are thus disconnected from the real router *a*. *e* will have a choice: the path calculated to *a* via *b* is of the same length as the path via *X*, as illustrated in figure 22(b).

177

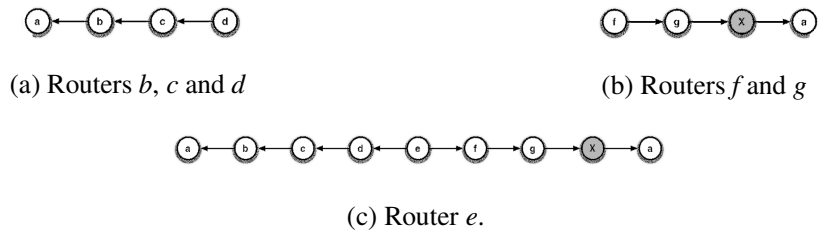

(a) Routers *b*, *c* and *d*                    (b) Routers *f* and *g*

(c) Router *e*.

Figure 22: Routing paths towards router *a*, as calculated by the different routers in the network

In general, the following observations can be made:

• The network will be separated in two, with those routers closer to *b* than to *X* reaching *a*, whereas those routers closer to *X* than to *b* unable to reach *a*.

• Routers beyond *b* will be unable to detect this link spoofing.

Returning to figure 19, if *X* advertises spoofed links to *a*, *b*, *c*, *d*, *f*, *g*, *h* and *i* in Link State Advertisements, the risk of detection by *e* is identical to if these were advertised in the Neighborhood Discovery process: *e* may detect that *X* is advertising links to *d* and *f*, while *X* is not recorded as a 2-hop neighbor via neither *d* nor *f*.

Suppressing links to *d* and *f* from being advertised by *X* would prevent *e* from detecting that *X* is malicious. However, upon receiving a Link State Advertisement, a router is able to detect if it itself is being spoofed – the advertising router is not a neighbor of the router being spoofed. Furthermore, for the reasons elaborated above, routers up to one hop away from the spoofed destination may detect the spoofing. In the case of figure 20, *d* would be able to detect spoofing of links to *c* (as would *c* be able to detect spoofed links to *b* etc.) – possibly leading to a significant fraction of routers being able to detect that *X* is conducting a disruptive attack and, therefore, engaging appropriate countermeasures. *e* would, in this case, be the only router unable to detect the spoofing. While this may suffice to disrupt the network, it is no different from the identity spoofing attack illustrated in figure 14, which carries less risk of detection of the malicious router.

The impact of this attack is similar to that presented in section 5.1.2, however, is easier to detect as the malicious router is generating control traffic reaching the entire network.

## 5.3  Creating Loops

Consider the example in figure 23(a). The malicious router, *X*, spoofs the identity of *g*, and participates (with this spoofed identity) in both the Neighborhood Discovery process and the Link State Advertisement process. In order to cover all its 2-hop neighbors, *a* must select both *X* and *c* as MPRs. Hence, the link *c-a* is advertised by *c*, and the link *g-a* is advertised by router *X*.

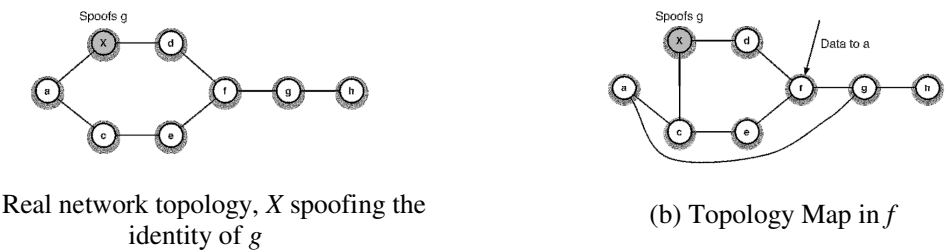

(a) Real network topology, *X* spoofing the          (b) Topology Map in *f*
identity of *g*

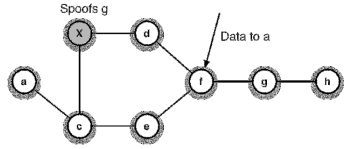

(c) Topology Map in *g*

Figure 23: Perceived Topology Maps identity spoofing in the Link State Advertisement process

The topology perceived by *f* is as indicated in figure 23(b): paths to the destination *a* exist via *g* (2 hops) or via *e* (3 hops). The topology perceived by *g* is as indicated in figure 23(c): as *g* does not process TCs originating from itself, the only path recognized by *g* towards *a* is via *f*. Therefore, if a data packet destined for *a* arrives at *f*, it will be forwarded through *g*. *g* will forward the data packet through *f*, thereby creating a loop in the network.

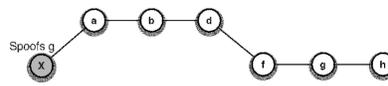

(a) Real network topology, with malicious router *X* spoofing the identity of *g*

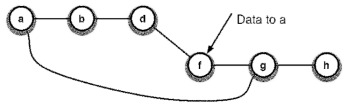

(b) Topology Map in router *f*

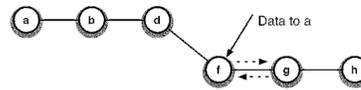

(c) Topology Map in router *g*

Figure 24: Perceived Topology Maps with identity spoofing in the Neighborhood Discovery

Consider the example in figure 24(a). The malicious router, *X*, spoofs the identity of *g*, and selects *a* as MPR. Hence, the link *a-g* is advertised by *a* – *a* is "tricked" into advertising a non-existing link. The topology perceived by *f* is as indicated in figure 24(b): paths to the destination *a* exists is via *g* (2 hops) or via *f* (3 hops). The topology perceived by *g* is as indicated in figure 24(c): as *g* does not process TCs originating from itself, the only path recognized by *g* towards *a* is via *f*. Therefore, if a data packet destined for *a* arrives at *f*, it will be forwarded through *g*. *g* will forward the data packet through *f*, thereby creating a loop in the network.

## 6 WHY THIS PAPER DOES NOT CONSIDER REPLAY ATTACKS

A commonly considered "attack" type is for a malicious router to record control traffic from legitimate routers, and "replay" this – possibly somewhere else in the network, and possibly at some later point in time. While such indeed is possible, it should not be considered as a class of attacks on OLSRv2 in and by itself: *In-fine*, the malicious router replaying messages is performing a combination of *identity-spoofing*, spoofing the identity of the router from which it recorded the messages, and *link-spoofing*, spoofing links to the (original) neighbors of that router. Thus, the impact of such a "replay attack" is no different from the impact described for identity-spoofing and link-spoofing.

## 7 INHERENT OLSRv2 RESILIENCE

While OLSRv2 does not specifically include security features (such as encryption), the protocol and its algorithms present some inherent resilience against part of the attacks described in this paper. In particular, it provides the following resilience:

179

• *Sequence numbers:* OLSRv2 employs message sequence numbers, specific per router identity and message type. Routers keep an "information freshness" number (ANSN), incremented each time the content of a Link State Advertisement from a router changes. This allows rejecting "old" information and duplicate messages, and provides some protection against "message replay". This, however, also presents an attack vector (section 4.3).

• *Ignoring uni-directional links:* The Neighborhood Discovery process detects and admits only bi-directional links for use in MPR selection and Link State Advertisement. Jamming attacks (section 3.2) may affect only *reception* of control traffic, however OLSRv2 will correctly recognize, and ignore, such a link as not bi-directional.

• *Message interval bounds:* The frequency of control messages, with minimum intervals imposed for HELLO and TCs. This limits the impact from an indirect jamming attack (section 4.5).

• *Additional reasons for rejecting control messages:* The OLSRv2 specification includes a list of reasons, for which an incoming control message should be rejected – and allows that a protocol extension may recognize additional reasons for OLSRv2 to consider a message malformed. This allows – together with the flexible message format [2] – addition of security mechanisms, such as digital signatures, while remaining compliant with the OLSRv2 specification.

## 8 CONCLUSION

This paper has presented a detailed analysis of security threats to the Optimized Link State Routing Protocol version 2 (OLSRv2), by taking an abstract look at the algorithms and message exchanges that constitute the protocol, and for each protocol element identifying the possible vulnerabilities and how these can be exploited. In particular, as link-state protocol, OLSRv2 assumes that (i) each router can acquire and maintain a topology map, accurately reflecting the effective network topology; and (ii) that the network converges, i.e. that all routers in the network will have sufficiently identical (consistent) topology maps. An OLSRv2 network can be effectively disrupted by breaking either of these assumptions, specifically (a) routers may be prevented from acquiring a topology map of the network; (b) routers may acquire a topology map, which does not reflect the effective network topology; and (c) two or more routers may acquire substantially inconsistent topology maps.

The disruptive attacks to OLSRv2, presented in this paper, are classified in either of these categories. For each, it is demonstrated if OLSRv2 has an inherent protection against the attack.

**Authors**

**Ulrich Herberg** graduated in computer science from TU Munich in 2007, and then joined Ecole Polytechnique (France), where he is currently a PhD student. His research interests cover routing and security issues in MANETs, auto-configuration, and delay-tolerant networks.

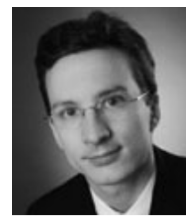

**Thomas Clausen** is a graduate of Aalborg University, Denmark (M.Sc., PhD - civilingeniør, cand. polyt), Thomas spent a number of years at INRIA where he, among other things, spent his time developing and standardising OLSR. In 2004 he joined faculty at Ecole Polytechnique, France's premiere technical university

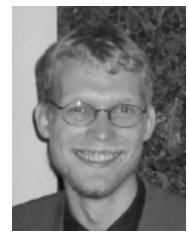